\documentstyle[aps,pra,amsfonts,psfig,twocolumn]{revtex}
\newcommand{\beq}{\begin{equation}}
\newcommand{\eeq}{\end{equation}}
\begin{document}
\title{Thermodynamics of photons in relativistic $e^+e^-\gamma$ plasmas}
\author{M.V. Medvedev\cite{mm}}
\address{Harvard-Smithsonian Center for Astrophysics,
60 Garden St., Cambridge, MA 02138}
\maketitle
\draft

\begin{abstract}
Thermodynamic and spectral properties of a photon gas in $e^+e^-\gamma$ plasmas 
are studied. The effect of a finite effective mass of a photon, associated with 
the plasma frequency cutoff, is self-consistently included.
In the ultra-relativistic plasma, the photon spectrum turns out to be 
universal with the temperature normalized plasma frequency cutoff being a 
fundamental constant independent of plasma parameters. Such a universality 
does not hold in the non-relativistic plasma.
\end{abstract}

It is well known that a photon gas which is in thermodynamic equilibrium with 
matter exhibits the Planckian blackbody energy spectrum. This is correct
only if matter is electrically neutral (for example, an atomic gas) where
the thermodynamic equilibrium is established via photon emission and absorption
in atomic transitions. Such an equilibrium no longer forms in ionized gases
(plasmas) where the number of charged particles is not negligible. 
An electromagnetic wave (photon) is strongly 
coupled to collective plasma excitations via oscillating electric and magnetic
fields. The dispersion relation of photons in a plasma is
$\omega^2\sim\omega^2_{\rm p}+k^2c^2$, where $\omega_{\rm p}$ is 
referred to as the plasma frequency which depends on
plasma parameters (e.g., densities of electrons and positrons). Quite 
interestingly, such a dispersion may be interpreted as a manifestation of a 
finite {\em effective mass} of a photon, $m_\gamma\sim\hbar\omega_{\rm p}/c^2$.
There are no photons with energies lower than its effective rest energy.
Hence, one may expect a cutoff in the Planckian spectrum for frequencies
$\omega<\omega_{\rm p}$. A new, modified blackbody, distribution of photons
forms in this case. The shape of the spectrum is controlled by the densities of 
electrons and positrons. It is remarkable that in the relativistic plasma, 
the electron and positron number densities are, in turn, 
{\em self-consistently} determined from the {\em equilibrium} 
between the photons and the electrons and positrons 
with respect to the $e^+e^-$-pair production.

Objects with hot, positron-electron-photon ($e^+e^-\gamma$) plasmas, referred
to as fireballs, are not rare in astrophysics. For example, such fireball
conditions are believed to be realized at early stages of the formation of the
Universe (see, for example, \cite{ZN}), in energetic cosmological explosions 
which produce $\gamma$-ray bursts (see \cite{Piran} for review), etc.. 
Hence, the study of thermodynamic and statistical properties
of a gas of photons in the $e^+e^-\gamma$ fireballs is of both academic and
practical interest. 

In this paper, we find a self-consistent, blackbody radiation spectrum and
thermodynamic properties of the photon gas in the $e^+e^-\gamma$ fireball
plasma. In particular, the equilibrium spectrum of photons in the
ultra-relativistic case ($T\gg m_ec^2$) turns out to be {\em universal}:
\begin{enumerate}
\item
the frequency cutoff normalized by temperature is a {\em fundamental constant}
independent of any plasma parameters; 
\item
the simple, power-law scalings of 
thermodynamic parameters such as pressure, density, etc. vs. temperature, valid 
for the Planck distribution, hold for ``massive'' photons as well. Numerical 
pre-factors, however, modify self-consistently. 
\end{enumerate}
In the end of the paper, 
we also briefly consider a non-relativistic case and discuss
implications for the cosmic microwave background spectrum distortion, purely 
radiative fireball model of $\gamma$-ray bursts , and stellar interiors. 

Here we should comment on one important point. The gas of photons obeys the
Bose-Einstein statistics with a vanishing chemical potential and, also, is
assumed to be in thermodynamic equilibrium with matter which, in general,
has a non-vanishing chemical potential. On the other hand, one of the
conditions of thermodynamic equilibrium is the equality of the chemical
potentials of the interacting systems. This contradiction may be resolved by
treating the process of radiation as a first-order phase transition (Bose
condensation) and properly taking into account the equality of the chemical
potentials of matter and radiation. An excellent discussion of this subject 
may be found in Ref.\ \cite{Lavenda-book}. The deviation of a photon spectrum
from the Planck distribution may be significant in the non-relativistic
plasmas, where thermodynamic equilibrium is established via, for example,
Compton scattering of a photon on $e^+$ or $e^-$. In the ultra-relativistic
regime (considered in this paper) two- and three- annihilation/creation of an
$e^+e^-$-pair dominate and are responsible for establishing thermodynamic
equilibrium. The chemical potential of an $e^+e^-$-pair is identically zero
(by definition, $e^+$ and $e^-$ have equal but opposite in sign chemical 
potentials) and the condition of the equality of the chemical potentials of
matter and radiation is satisfied automatically. The Planck statistics of
photons is thus valid in the ultra-relativistic $e^+e^-\gamma$ plasma. 

{\em Ultra-relativistic case} --- The dispersion relation of electromagnetic
waves in the ultra-relativistic plasma reads \cite{Silin}:
\beq
\frac{k^2c^2}{\omega^2}=1+\frac{3}{4}\frac{\omega_{\rm p,rel}^2}{\omega kc}
\left[\left(1-\frac{\omega^2}{k^2c^2}\right)
\ln{\left|\frac{\omega-kc}{\omega+kc}\right|}-\frac{2\omega}{kc}\right] ,
\label{disp}
\eeq
where $\omega_{\rm p,rel}^2=\sum_\alpha 4\pi e_\alpha^2n_\alpha c^2/3T$ is
the relativistic plasma frequency, $e_\alpha$ and $n_\alpha$ are the charge and
number density of species $\alpha$, the sum, $\sim_\alpha$, goes over all 
relativistic ($T\gg m_\alpha c^2$) species, i.e., electrons and positrons in 
our case, and the contribution from non-relativistic species being $mc^2/T$ 
times smaller is neglected here. An exact analytical solution to this equation 
is unknown. The asymptotic behavior
\beq
\omega^2=\left\{
\begin{array}{ll}
\displaystyle{\omega_{\rm p,rel}^2+\frac{5}{6}k^2c^2}, & \omega\gg kc;\\
\displaystyle{\frac{3}{2}\omega_{\rm p,rel}^2+k^2c^2}, & \omega\to kc,
\end{array}
\right.
\eeq
suggests, however, the following approximate solution to Eq.\ (\ref{disp}):
\beq
\omega^2=\omega_{\rm p,rel}^2+k^2c^2,
\eqnum{$1'$}
\label{approx}
\eeq
which roughly satisfies both asymptotics and, also, is the {\em exact} 
dispersion relation in the non-relativistic plasma. A direct numerical 
comparison of Eqs.\ (\ref{disp}) and (\ref{approx}) shows that the fractional 
error of the approximation does not exceed 3\%.

In the ultra-relativistic plasma, the electrons and positrons are in 
thermal equilibrium. Hence, their number densities are equal and 
$n_{e^+}=n_{e^-}=\frac{7}{8}n_\gamma$, where the factor $7/8$ accounts for
the difference in Fermi and Bose statistics \cite{Kolb-Turner} and
$n_\gamma$ is the number density of photons. 
The relativistic plasma frequency becomes now
\beq
\omega_{\rm p,rel}^2=\sum_\pm\frac{4\pi e^2n_\pm c^2}{3T}
=\frac{7\pi e^2n_\gamma c^2}{3T} .
\label{omega}
\eeq
Note here that the plasma frequency depends on the density of photons 
to be determined \cite{comment}. 

The occupation numbers for a gas of photons are given by the Bose-Einstein
distribution with a vanishing chemical potential: ${\rm n}(k)=\left[
\exp{\left(\hbar\omega(k)/T\right)}-1\right]^{-1}$.
The number density of photons is formally written as follows
\cite{Landafshits}:
\beq
n_\gamma=\frac{T^3}{\pi^2\hbar^3c^3}\int_0^\infty\frac{x^2{\rm d}x}
{\exp\left(\sqrt{7\pi e^2\hbar^2c^2n_\gamma/3T^3+x^2}\right)-1} 
\label{n}
\eeq
where we made use of the fact that the degeneracy $g=2$ for photons (two
independent polarizations). This equation {\em self-consistently} determines
the number density of photons in the $e^+e^-\gamma$ fireball plasma where
the plasma frequency cutoff (i.e., the effective photon mass), 
in turn, depends on $n_\gamma$. 

To proceed further we note a remarkable fact. Namely, $n_\gamma$ and $T$ enter
Eq.\ (\ref{n}) in combination $n_\gamma/T^3$. Then, the solution to this
equation must be sought in the form $n_\gamma=const.\times T^3$ with the
$const.$ to be determined from Eq.\ (\ref{n}). Thus, the parameterization 
$n_\gamma\propto T^3$ removes {\em explicit} temperature
dependence of the integral in Eq.\ (\ref{n})! Hence this solution for
$n_\gamma$ is {\em universal}. Note, this is
usually not the case for particles with nonzero mass. Since ${\rm n}({\bf p})=
\left[\exp\left(\sqrt{m^2c^4+|{\bf p}|^2c^2}/T\right)-1\right]^{-1}$, 
temperature enters the integral over particle momenta, ${\bf p}$, 
via the combination $(mc^2/T)^2$, so that all the thermodynamic quantities 
(e.g., density, energy, entropy, etc.) do not exhibit universal, 
power-law scalings. For photons, however, $m_\gamma\propto T$ and the
universal (Planckian) scalings are recovered. Numerical coefficients, 
however, modify self-consistently.

Now, we write $n_\gamma=f\, n_{\rm Pl}$, where 
$n_{\rm Pl}=\left[2\zeta(3)/\pi^2\right]\left(T/\hbar c\right)^3$ and 
$\zeta(3)\simeq1.2$ is the Riemann zeta function of 3 and $f$ is a constant
to be calculated. Hereafter the subscript ``Pl'' denotes values calculated 
for the Planckian distribution. Then Eq.\ (\ref{n}) becomes
\beq
2\zeta(3)f=\int_0^\infty\frac{x^2{\rm d}x}
{\exp\left(\sqrt{\Delta^2f+x^2}\right)-1} ,
\label{f}
\eeq
where 
\beq
\Delta=\frac{\hbar\omega_{\rm p,rel}}{T}
=\left(\alpha_e\frac{14\zeta(3)}{3\pi^2}\right)^{1/2}\simeq6.44\times10^{-2} 
\label{Delta}
\eeq
and $\alpha_e=e^2/\hbar c$ is the fine structure constant. Given $f$, 
the energy density, pressure of the gas of photons as well as other 
thermodynamic parameters may be calculated straightforwardly 
\cite{Landafshits}:
\begin{mathletters}
\begin{eqnarray}
\epsilon_\gamma&=&\frac{T^4}{\pi^2\hbar^3c^3}\int_0^\infty
\frac{\sqrt{\Delta^2f+x^2}\;x^2{\rm d}x}
{\exp\left(\sqrt{\Delta^2f+x^2}\right)-1} ,
\label{e}\\
p_\gamma&=&\frac{1}{3}\frac{T^4}{\pi^2\hbar^3c^3}\int_0^\infty
\frac{x^3{\rm d}x}
{\exp\left(\sqrt{\Delta^2f+x^2}\right)-1} .
\end{eqnarray}
\label{e-p}
\end{mathletters}
Numerically solving Eq.\ (\ref{f}) for $f\equiv(1+\delta f_n)$ and using 
Eqs.\ (\ref{e-p}), one finally
obtains:
\begin{mathletters}
\begin{eqnarray}
&n_\gamma=(1+\delta f_n) n_{\rm Pl}, 
	&\qquad\delta f_n\simeq3.4\times10^{-3} ;\label{correction-n}\\
&\epsilon_\gamma=(1+\delta f_\epsilon) \epsilon_{\rm Pl}, 
	&\qquad\delta f_\epsilon\simeq5.2\times10^{-4} ;\\
&p_\gamma=(1+\delta f_p) p_{\rm Pl}, 
	&\qquad\delta f_p\simeq1.0\times10^{-3} ,
\end{eqnarray}
\label{corrections}
\end{mathletters}
where $\epsilon_{\rm Pl}=3p_{\rm Pl}=(\pi^2/15)(T^4/\hbar^3c^3)$.
It is well known that the equation of state of a gas of non-relativistic 
massive particles, $T\ll mc^2$, is $p_{\rm nr}=\frac{2}{3}\epsilon_{\rm nr}$
and that of ultra-relativistic (or massless) particles, $T\gg mc^2$, is
$p_{\rm r}=\frac{1}{3}\epsilon_{\rm r}$. Since photons in the relativistic 
plasma turn out to be ``sub-relativistic'', 
$\hbar\omega_{\rm p,rel}/T\sim5\times10^{-2}$, we expect that the effective 
equation of state of photons will be 
$p_\gamma=a\epsilon_\gamma,\ {\frac{1}{3}<a<\frac{2}{3}}$. Indeed, from
Eqs.\ (\ref{corrections}) it follows $p_\gamma/\epsilon_\gamma\simeq
(1+\delta f_p-\delta f_\epsilon)p_{\rm Pl}/\epsilon_{\rm Pl}$, and we have:
\beq
p_\gamma\simeq\frac{1}{3}(1+5.0\times10^{-4})\epsilon_\gamma .
\eeq

\begin{figure}
\psfig{file=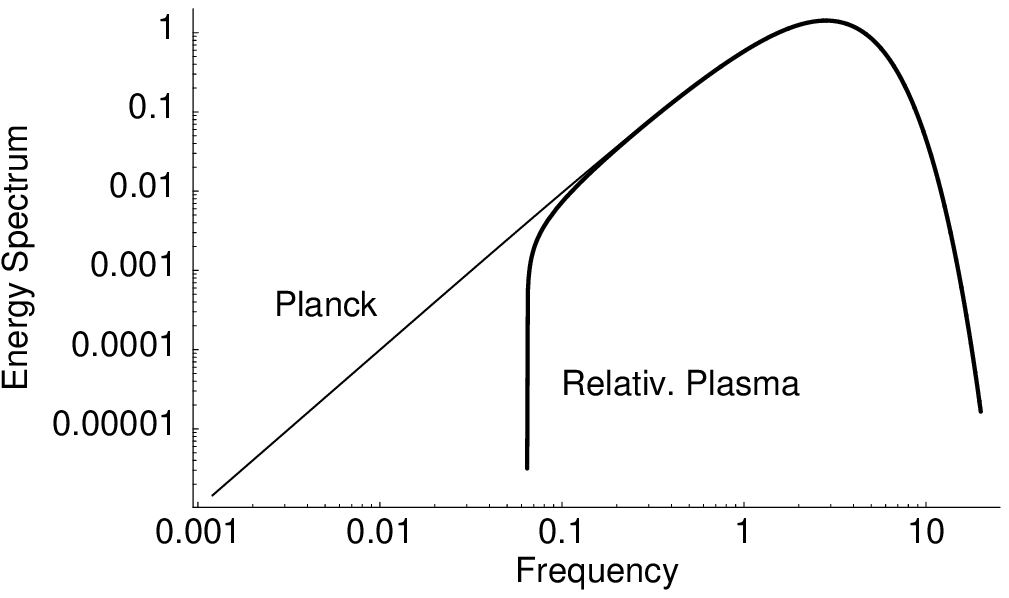,width=3.5in}
\caption{The normalized blackbody equilibrium spectrum of radiation from 
the $e^+e^-\gamma$ relativistic plasma and the Planckian spectrum vs. 
dimensionless frequency, $\omega_*=\hbar\omega/T$.}
\label{fig-spectra}
\end{figure}

The blackbody equilibrium radiation spectrum from the ultra-relativistic 
$e^+e^-\gamma$ fireball plasma is easily obtained from Eq.\ (\ref{e}):
\beq
{\rm d}\epsilon_\gamma(\omega_*)=\frac{T^4}{\pi^2\hbar^3c^3}
\frac{\omega_*^2\sqrt{\omega_*^2-\Delta^2f}}{e^{\omega_*}-1}\ {\rm d}\omega_* ,
\eeq
where $\omega_*=\hbar\omega/T$ is the dimensionless frequency.
Note again that this spectrum is {\em universal}. The low-frequency cutoff
scales with temperature in such a manner that 
$\hbar\omega_{\rm p,rel}/T=\Delta\sqrt{f}=const.$. 
The value of the low-frequency cutoff is a {\em fundamental constant}
given by Eq.\ (\ref{Delta}) [correction due to $f$, Eq.\ (\ref{correction-n})
is small and may be neglected] and independent of any parameters of the plasma.
The universal, blackbody energy spectrum of radiation from the $e^+e^-\gamma$
relativistic plasma is shown in Fig.\ \ref{fig-spectra}. The Planckian
spectrum is shown for comparison.

At last we comment that the radiation from a relativistic fireball is 
{\em always blackbody}. Indeed, for a non-thermal radiation spectrum to 
occur, the linear size of the system should be 
$R<\tau/\sigma_{\rm T}n_{\rm Pl}(T\sim m_ec^2)\lesssim10^{-4}~\textrm{cm}$
($\sigma_{\rm T}$ is the Thompson cross-section and $\tau\lesssim1$ is the
optical depth of the plasma) which is microscopically small.

{\em Non-relativistic case} --- The universality of the photon spectrum breaks
down at plasma temperatures below $T\sim m_ec^2$. The dispersion relation is 
exactly Eq.\ (\ref{approx}) with $\omega^2_{\rm p,rel}$ replaced by its
non-relativistic counterpart:
\beq
\omega_{\rm p}^2=\frac{4\pi e^2n_{e^\pm}}{m_e} ,
\eeq
where $n_{e^\pm}\equiv n_{e^+}+n_{e^-}$ and contributions from heavier
particles are neglected. When the temperature drops below $m_ec^2$ the number
density of $e^+e^-$-pairs decreases exponentially due to annihilation.
Hence, we expect the non-relativistic cutoff frequency to be smaller than the
ultra-relativistic. The density of charged particles is easily calculated
\cite{Landafshits}:
\beq
n_{e^\pm}=\left[n_0^2+\frac{3}{\pi^3}\left(\frac{m_ec}{\hbar}\right)^6\!
\left(\frac{T}{m_ec^2}\right)^3\!\!
\exp\!\!\left(-\frac{2m_ec^2}{T}\right)\right]^{-1/2} .
\eeq
where $n_0$ is the residual density of electrons at $T\to0$ (i.e., in absence
of pair creation).

A few estimates below show that the effect of the plasma frequency cutoff is
usually small, unless a plasma is extremely dense and `cold'.
First, most of the explosion energy in cosmological, purely radiative
fireballs (if any), --- the $\gamma$-ray bursters, --- is emitted 
when the system becomes optically thin for radiation. This occurs at
$T\simeq20~\textrm{keV}$ \cite{Piran} when the $e^+e^-$-pair density drops to 
$n_{e^\pm}\sim1/\sigma_{\rm T}R_{\rm thin}\sim10^{14}~\textrm{cm}^{-3}$. 
The low-frequency cutoff is 
$\left(\hbar\omega_{\rm p}/T\right)_{\rm GRB}\sim10^{-8}$ 
which is hardly observable.
Second, distortions of the cosmic microwave background energy spectrum 
may be traced up to redshifts of $z\sim10^8$ \cite{ZN}. Taking the residual
density of electrons $n_0\sim10^{-9}n_{\rm Pl}$ and temperature 
$T\sim\textrm{few keV}$, we estimate the cutoff 
$\left(\hbar\omega_{\rm p}/T\right)_{\rm CMB}\sim10^{-7}$.
Third, in the center of the Sun $n_0\sim10^{26}~\textrm{cm}^{-3}$ and 
$T\sim1.4~\textrm{keV}$ which yields the cutoff
$\left(\hbar\omega_{\rm p}/T\right)_{\rm Sun}\sim0.2$; that is the cutoff 
is close to the Planckian thermal peak. The effect on the solar (and stellar)
models will be addressed in future publications.

To conclude, the thermodynamics of a gas of photons in the $e^+e^-$ plasma
is studied. The deviation from the Planck distribution occurs because no
photons with frequencies below the plasma frequency exist in the plasma.
Equivalently, photons in the plasma acquire an effective mass.
In the ultra-relativistic plasma, $T\gg m_ec^2$, the effective 
photon mass scales as $m_\gamma=T\Delta/c^2$ with $\Delta$ being the
fundamental constant (independent of any plasma parameters) given by Eq.\ 
(\ref{Delta}). Hence, the equilibrium (blackbody) spectrum exhibits universal
properties. The power-law scalings of thermodynamic parameters vs. temperature
for ``massive'' photons are the same as those for the Planck distribution, but
with different numerical pre-factors. In contrast, thermodynamics of photons 
in the non-relativistic plasma is not universal.

The author is grateful to Ramesh Narayan and Abraham Loeb for useful and 
stimulating discussions.

\begin{references}
\bibitem{comment} If there are other ultra-relativistic (i.e., effectively
massless), charged species, then $n_\pm=g_*'g_\gamma n_\gamma$ in Eq.\
(\ref{omega}), where $g_\gamma=2$,\ 
$g_*'=\sum_{i={\rm bosons}}'g_i+\frac{7}{8}\sum_{i={\rm fermions}}'g_i$, and
`prime' denotes summation over charged, ultra-relativistic species only.
\end{references}
\end{document}